\documentclass{llncs}
\usepackage{graphicx}
\graphicspath{ {images/} }
\usepackage{multirow,ifsym}
\usepackage{caption}
\usepackage{bbding}
\usepackage{subfigure}
\usepackage[table]{xcolor}
\usepackage[labelfont=bf]{caption}
\begin{document}

\title{A Radiomics Approach to Computer-Aided Diagnosis with Cardiac Cine-MRI}
\author{Irem Cetin\inst{1} \Envelope , Gerard Sanroma\inst{1}, Steffen E. Petersen\inst{3}, Sandy Napel\inst{4}, 
\\Oscar Camara\inst{1}, Miguel-Angel Gonzalez Ballester\inst{1,2}, Karim Lekadir\inst{1}}
\institute{Universitat Pompeu Fabra, BCN MedTech, Barcelona, Spain {\email{irem.cetin01@estudiant.upf.edu}}
\and
Catalan Institution for Research and Advanced Studies (ICREA), Barcelona, Spain
\and
Queen Mary University of London, William Harvey Research Institute, London, UK
\and
Stanford University, School of Medicine, Department of Radiology, Stanford, USA}
\maketitle
\begin{abstract}
Computer-aided diagnosis of cardiovascular diseases (CVDs) with cine-MRI is an important research topic to enable improved stratification of CVD patients. However, current approaches that use expert visualization or conventional clinical indices can lack accuracy for borderline classications. Advanced statistical approaches based on eigen-decomposition have been mostly concerned with shape and motion indices. In this paper, we present a new approach to identify CVDs from cine-MRI by estimating large pools of radiomic features (statistical, shape and textural features) encoding relevant changes in anatomical and image characteristics due to CVDs. The calculated cine-MRI radiomic features are assessed using sequential forward feature selection to identify the most relevant ones for given CVD classes (e.g. myocardial infarction, cardiomyopathy, abnormal right ventricle). Finally, advanced machine learning is applied to suitably integrate the selected radiomics for final multi-feature classification based on Support Vector Machines (SVMs). The proposed technique was trained and cross-validated using 100 cine-MRI cases corresponding to five different cardiac classes from the ACDC MICCAI 2017 challenge \footnote{https://www.creatis.insa-lyon.fr/Challenge/acdc/index.html}. All cases were correctly classified in this preliminary study, indicating potential of using large-scale radiomics for MRI-based diagnosis of CVDs.
\end{abstract}
\section{Introduction}\label{sec:Introduction}
Despite continuous progresses both in clinical research and practice, cardiovascular diseases (CVDs) remain the leading cause of mortality and morbidity in the world \cite{1}. In this context, cardiac imaging such as cine-MRI (Magnetic Resonance Imaging) is expected to play an important role due to its ability to quantify in detail structural and functional properties of the beating heart \cite{2}. However, visual assessment of CVDs using cine-MRI remains challenging and labor-intensive due to the complexity of these diseases, in particular when the structural and functional disorders are subtle  \cite{3}. Moreover, quantitative assessment through existing clinical indices such as volumetric measures, ejection fraction, and thickening measures can be suboptimal for borderline cases. Consequently, more advanced automated techniques are needed to exploit the richness of the cardiac data to estimate diagnosis, as well as severity of the phenotype which often is associated with prognosis. Over the years, several methods have been proposed based on eigendecomposition of the moving cardiac shapes \cite{4,5,6,7,8,9,10} but they have mostly used geometrical information.

In this paper, we propose instead a radiomics approach to automated image-based diagnosis of complex CVDs. Radiomics is the task of calculating a large number of imaging descriptors from delineated images, which has been widely used in cancer imaging \cite{11}. In the context of cine-MRI, radiomics describing changes in image appearence due to CVDs have not been exploited. The proposed method estimates a large number of radiomic features including statistical, shape, and textural descriptors and assess their ability to discriminate between different CVDs automtically and robustly within a machine learning framework based on SVMs. Training and cross-validation are carried out in this study based on a database of 100 cine-MRI cases corrresponding to five different subclasses, from the ACDC challenge of MICCAI 2017.
\section{Method}\label{sec:Method}
\subsection{Data Description}\label{sec:Data}
\begin{figure}
\centering  
\subfigure[DCM]{\includegraphics[width=0.22\textwidth]{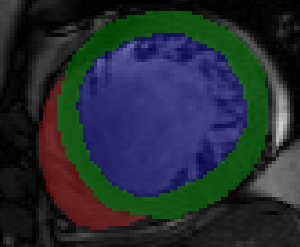}}
\subfigure[HCM]{\includegraphics[width=0.22\textwidth]{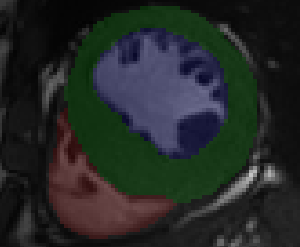}}
\subfigure[MINF]{\includegraphics[width=0.22\textwidth]{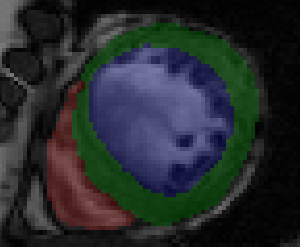}}
\subfigure[RV]{\includegraphics[width=0.22\textwidth]{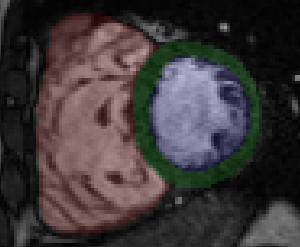}}
\centering  
\subfigure{\includegraphics[width=0.22\textwidth]{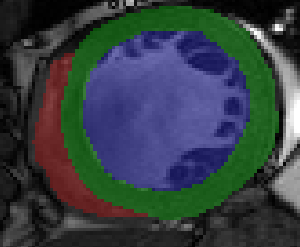}}
\subfigure{\includegraphics[width=0.22\textwidth]{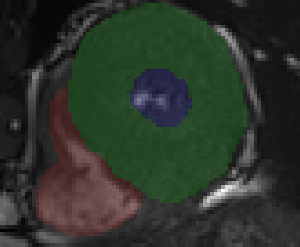}}
\subfigure{\includegraphics[width=0.22\textwidth]{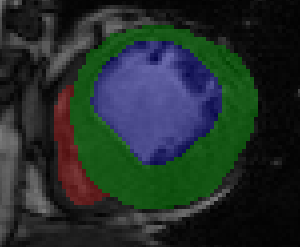}}
\subfigure{\includegraphics[width=0.22\textwidth]{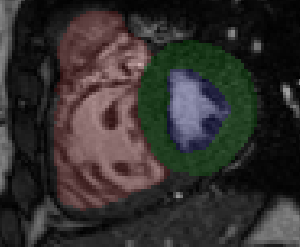}}
\caption{Examples of cine-MRI images for the four abnormalities classified in this study (Top: ED, bottom: ES).}
\label{fig:data}
\end{figure}

This study was conducted in the context of the MICCAI 2017 challenge on Automated Cardiac Diagnosis (the ACDC challenge). The database consists of 100 cases comprising cine-MRI data, height and weight information, as well as the diastolic and systolic phase instants for each subject. Five subclasses were included, namely (see examples in Fig. \ref{fig:data}):
\\(1) Normal subjects (NOR).
\\(2) Patients with dilated cardiomyopathy (DCM).
\\(3) Hypertrophic cardiomyopathy (HCM). 
\\(4) Abnormal right ventricle (RV)
\\(5) Myocardial infarction (MINF). 

The 100 images were acquired at the University Hospital of Dijon (France) by using  1.5T or 3T MRI scans (Siemens Medical Solutions, Germany) with the following parameters depending on the examination:  image sequence = SSFP cine-MRI, slice thickness = 5mm or 8mm, inter-slice gaps = 5mm or 10mm, spatial resolution = 1.37 to 1.68 mm2/pixel, number of frames = 28 to 40.  This training dataset was then manually segmented for the left ventricle, myocardium and right ventricle by an experienced manual observer at both end-diastole and end-systole time frames. The test data of the challenge will be segmented as explained in the next section.

\subsection{Semi-Automatic Segmentation}\label{sec:Radio}
To segment the test datasets semi-automatically, we propose an atlas-based approach using the publicly available cardiac atlas \cite{12}. To this end, we firstly define manually six anatomical landmarks on each cine-MRI case, more specifically at:
\\ (1) Mid-ventricular slice: RV insertion point next to the liver.
\\ (2) Mid-ventricular slice: A point on the RV free wall.
\\ (3) Mid-ventricular slice: RV insertion point next to the lung.
\\ (4) Mid-ventricular slice: A point on the LV free wall.
\\ (5) Apical slice: Apex.
\\ (6) Basal slice: Center of the base. 

We then use the atlas-based technique described in \cite{13} to extract the cardiac structures of interest, namely the LV, RV and myocardium. This is followed by user-friendly manual correction of the segmented contours to correct for potential errors using the ITK-SNAP tool\footnote{http://www.itksnap.org/pmwiki/pmwiki.php}. Note that this segmentation approach will be only used to segment the test data of the challenge and that this paper focuses only on the classification part of the ACDC challenge.
\subsection{Radiomics Features for Cardiac Diagnosis}\label{sec:Radio}
As mentioned in the introduction, most existing techniques included in clinical practice use shape and motion indices such as ejection fraction, ventricular volumes and myocardial thickening to classify the subjects under investigation. This means that a lot of information produced by the image is lost during this operation, in particular imaging evidence in relation to the tissue appearence in the blood pool, myocardium and right ventricle, as well as more complex morphological and functional information. But it is unclear which advanced indices could contribute to improved classification of cardiovascular cases. To address these issues, we propose a radiomics approach for computer-aided diagosis in cine-MRI. Radiomic features have been used so far mostly for cancer image quantification \cite{11,13}, such as for the estimation of patient prognosis and treatment response based on the characteristics of the tumors as encoded by the image data. The suffix -omics attached to radiology refers to the use of large amounts of imaging features, from which the most relevant set can be selected for the specific task in question. In this paper, we estimate a large pool of new radiomic features from the segmented cine-MRI images, which will be then analyzed to extract the most powerful features for classification.

In other words, we augment the set of indices to be leveraged for cardiac diagnosis by considering more complex shape/motion radiomic features, as well as advanced textural radiomic features. Specifically, we use 567 features (height, weight, ED-ES duration, plus 188 features per structure: LV, MYO, RV at ED and ES) based on five categories using the PyRadiomics library \cite{14}, namely:
\\(1) Shape based (Volume, surface area, sphericity, compactness, diameters, elongation, etc).
\\(2) Intensity first order statistics (e.g. mean, standard deviation, energy, entropy, etc).
\\(3) Gray level cooccurence matrix (GLCM) (autocorrelation, contrast, dissimilarity, homogeneity, inverse difference moment, maximum probability, etc).
\\(4) Gray level run length matrix (GLRLM) (short/long run emphasis, gray-level/run-length non-uniformity, etc).
\\(5) Gray level size zone matrix (GLSZM) (smal/large area emphasis, zone percentage, etc).

Note that the first group of radiomics consist of pure shape information, while the four remaining groups are intensity-based features, describing the intensity variations inside the cardiac structures, as well as the complexity and repeatability of the tissue texture. Our hypothesis is that some of these radiomic values will be modified in the presence of cardiac abnormality in a way that is unique to each subgroup of patients when compared to normal individuals.
\subsection{Classification Method}\label{sec:Class}
The next step of our approach is to combine the heterogeneous radiomic features within a classification scheme that will learn to discriminate between the different patient subgroups and normal individuals. In this paper, we choose to use Support Vector Machines (SVMs) \cite{15} due to well-known performance when classifying image data, in particular in the case of small sample size. An SVM model corresponds to a transformation of the examples to a hyperspace where a good separation is achieved by the hyperplanes that have the largest distance to the nearest training-data point of any class (so-called functional margin). This ensures that the examples belonging to the different classes are separated as clearly as possible. New cases are then mapped onto that same hyperspace and classified based on their location with respect to the hyperplanes separating the different classes. As such, it is suitable for cardiovascular disease classification as the challenge is precisely to identify subtle changes and differences between normal cardiac characteristics and those of pathological cases. 
\subsection{Radiomic Feature Selection}\label{sec:Select}
Due to the large number of radiomic features, radiomic-based classification can easily suffer from over-fitting due to the limited number of examples that can be realistically collected for training. As a result, it is of paramount importance to identify a smaller subset of radiomic features that is optimal for the cardiac diagnosis task. In this paper, we do this by using sequential forward feature selection \cite{16}, through which radiomic features will be added to the final subset one at a time until the classification becomes negatively impacted as a result of adding new radiomic features.
\section{Results}\label{sec:Results}
For all experiments, we used leave-one-out tests to evaluate the proposed method and measured accuracy as proportion correct classifications. Firstly, we evaluated the accuracy of the CVD classifications by using only intensity radiomics or only shape radiomic features. For intensity radiomics, we obtained a maximal accuracy of 0.98 (two misclassifications) when using 13 optimal features. For shape radiomics, we obtained an accuracy of 1.0 (all cases correctly classified) but by using a total of 32 features. Subsequently, we combined intensity, shape and patient information (height and weight) all together and the forward feature selection results are provided in Fig. \ref{fig:acc}. It can be seen that the best single feature only achieves a 0.62 accuracy. However, after adding three selected features to the classification task, the accuracy is improved beyond the 0.90 accuracy line to reach 0.91 and even 0.94 after five selected features. A maximum accuracy of 1.0 (all cases correctly classified) is reached by combining 10 features only when combining intensity, shape and patient information. 

\begin{center}
\includegraphics[scale=0.65]{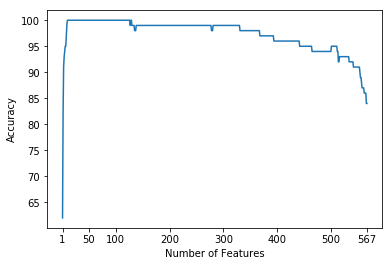}
\end{center}
\captionof{figure}{Accuracy of the proposed CVD classification as a function of the number of radiomic features trained in the model.}
\label{fig:acc}
\vspace{5mm} 

The shape of the curve in Fig. \ref{fig:acc} indicates the importance of feature selection, as after reaching the maximum accuracy, incorporating additionnal features leads to model over-fitting and reduced accuracy. While these preliminary results are obtained on a small and controlled study, they are encouraging. In comparison, we obtained an accuracy of 0.84 when using all radiomic features and 0.86 when combining conventional clinical indices only such as ejection fraction, cavity volumes, and body-mass index.

\setlength{\parindent}{4ex} 
To understand the behavior of the model, we evaluated the precision, recall and confusion matrix after selecting five features, at an accuracy of 0.94. It can be seen that the least accurately detected classes are for the HCM and DCM patients. However, after adding all optimal features, we finally reach a maximal accuracy of 1.0.

 \captionof{table}{Precision, recall and confusion matrix obtained by using the five first optimal radiomic features at accuracy of 0.94.} 
\begin{minipage}{0.5\textwidth}
    \begin{tabular}{|l |c|  c | c|  c|   c|   }\hline
       & NOR & DCM & HCM & MINF & RV\\ \hline
    Precision & 1 & 0.85 & 0.9 & 0.95 &1\\ 
    Recall    & 0.87 & 1 & 0.86 & 1 &1  \\
    \hline
    \end{tabular}
\end{minipage}
\begin{minipage}{0.5\textwidth}
    \begin{tabular}{|l|  c|  c | c |c | c |  }\hline
          & NOR & DCM & HCM & MINF & RV\\ \hline
    NOR & 20 & 0 & 0 & 0 &0\\ 
    DCM & 0 & 17 & 0 & 3 &0  \\ 
    HCM & 2 & 0 & 18 & 0 &0  \\ 
    MINF & 1 & 0 & 0 & 19 &0  \\ 
    RV  & 0 & 0 & 0 & 0 &20\\
    \hline
    \end{tabular}
\end{minipage}
   
\label{table:conf}
\vspace{5mm} 
This selected list of optimal features is given in Table \ref{table:feat}, which include one conventional shape index (volume), seven advanced shape radiomic features (e.g. compactness, least axis, surface area), one patient information (height) and one textural radiomic feature (GLCM inverse difference). This shows how mutiple radiomics of different nature can be complimentary to each other, which enables to identify correctly all cases.  Also, the table shows that the features are well distributed among the three cardiac structures (LV, myocardium, RV), as well as for the ED and ES frames.  

\setlength{\parindent}{4ex} 
To show the relevance of the selected features, we have added to the table the accuracy results by removing each feature from the SVM model (column W/O). It can be seen that the removal of each of these features negatively affects the final accuracy, which is reduced from 1.0 to 0.88 by removing the Surface Area to Volume feature, and to 0.96 by removing the Inverse Difference Intensity (GLCM) feature. This shows how these features can play an role in discriminating some of the challenging and ambiguous cases.

\setlength{\parindent}{4ex} 
To further show the relevance of combining all of the selected, we have also added to the table in the last column the accuracy by using a single radiomic feature. It can be seen that their on their own, these features do not enable a satisfactory classification, with the accuracy values vaying between 0.18 (Height) and 0.62 (Volume). In particular, the Height variable is not capable of producing any meaninful classification on its own, but contributes to the overall accuracy of the multi-radiomic model by normalizing with respect to size.   
 
\captionof{table}{List of 10 selected radiomic features as selected by the proposed technique for CVD classification. W/O: Accuracy without the feature. Alone: Accuracy using only this feature.}   
\begin{center}
 \begin{tabular}{|l|c|c|c|c|c|} \hline
    
    	  Name & Type & Frame & Structure &W/O & Alone \\ \hline
    	  Volume & Conventional shape & ED & MYO & 0.92 & 0.5\\ 
    	  Surface Area to Volume & Advanced shape & ES & LV & 0.88 & 0.62\\ 
    	  Least Axis & Advanced shape & ES & LV & 0.95 & 0.42\\ 
    	  Maximum 2D diameter & Advanced shape & ED & LV &  0.95 & 0.41\\ 
    	  Maximum 3D diameter & Advanced shape & ES & RV & 0.97 & 0.36\\ 
    	  GLCM Inverse Difference & Intensity/textural & ES & RV & 0.96 & 0.34\\ 
    	  Compactness 2 & Advanced shape & ES & LV  & 0.91 & 0.40\\
    	  Maximum 3D diameter & Advanced shape & ES & MYO & 0.96 & 0.47\\ 
    	  Surface area & Advanced shape & ED & RV & 0.97 & 0.29\\
    	  Height & Patient Information & - & - & 0.91 & 0.18\\ 
	\hline
	\end{tabular}

\label{table:feat}
 \end{center} 
 
\section{Conclusions}\label{sec:Conclusion}
In this paper, we proposed the use of large amounts of radiomic features, integrating advanced shape and textural descriptors, to predict cardiac subgroups. The obtained results suggest that radiomics are indeed capable to encode alterations in the anatomy and tissues of the affected cardiac structures. Furthermore, the feature selection results indicate that shape and intensity descriptors compliment each other and their combinations enable to enhance the prediction power of the system in particular for uncertain cases situated close to the boundary between two disease classes. However, the high accuracy of 1.0 suggests that further evaluations with additional datasets are required to test this radiomics model in larger and more variable data samples. In particular, inter-subject variability due to semi-automatic segmentation of the boundaries will need to be assessed. Future work also includes the testing of additional radiomic features (e.g. fractals, wavelets) and clincal interpretation of the features and results.

\section*{Acknowledgments}\label{sec:Acknowledgments}
IC and KL are funded by a Ramon y Cajal research grant (Ryc-2015-17183) from the Spanish Ministry of Economy and Competitiveness. SN is partly funded by a National Institute of Health grant (NIH U01 CA187947). The work of SEP forms part of the translational research portfolio of the NIHR Biomedical Research Unit at Barts.

\end{document}